\documentclass{article}
\def\p{\partial}
\def\s{\sigma}

\def\G{\Gamma}
\def\g{\gamma}

\def\de{\delta}

\def\ld{\lambda}
\def\Ld{\Lambda}

\def\ep{\epsilon}
\def\e{\eta}

\def\rh{\rho}

\def\b{\beta}

\def\ol{\overline}

\def\a{\alpha}

\def\pdellx'{\frac{\partial}{\partial x'}}
\def\pdellw'{\frac{\partial}{\partial w'}}

\newcommand{\be}{\begin{equation}}
\newcommand{\ee}{\end{equation}}
\def\bed{\begin{displaymath}}
\def\eed{\end{displaymath}}
\def\bea{\begin{eqnarray}}
\def\eea{\end{eqncrray}}
\def\[{$$}
\def\]{$$}

\begin{document}

\title{The S-matrix and graviton self-energy   \\  
in quantum Yang-Mills gravity} 
\author{ Jong-Ping Hsu\footnote{e-mail:jhsu@umassd.edu} and Sung Hoon Kim\footnote{e-mail:skim2@umassd.edu}  \\  
Department of Physics,
 University of Massachusetts Dartmouth \\
 North Dartmouth, MA 02747-2300, USA }


\maketitle
{\small  The S-matrix, its unitarity and the graviton self-energy at the one-loop level are discussed on the basis of  quantum  Yang-Mills gravity with the translational gauge 
symmetry in flat space-time.   The unitarity and gauge invariance of 
the S-matrix in a class of gauge conditions is preserved by massless ghost vector
particles, called `Feynman-DeWitt-Mandelstam' (FDM) ghosts, in quantum Yang-Mills gravity.  Using dimensional regularization, the graviton self-energy are explicitly calculated with a general gauge condition.  The resultant divergence of graviton self-energy at the one-loop level resembles to that in quantum electrodynamics.}

\bigskip
\bigskip
 
 \hspace{1.9in} {\small In memory of Ching-Chiang Chen, who }
 
 \hspace{1.9in} {\small introduced JP to Utiyama's exciting idea }
 
 \hspace{1.9in} {\small  about  gravity based on Yang-Mills gauge }
 
 \hspace{1.9in} {\small  symmetry during our college years.}

\bigskip


\bigskip

\section{Introduction}
\noindent

The idea of a gravitational theory with space-time translational gauge symmetry is very interesting from the viewpoint of Yang-Mills theory.    It has attracted many authors \cite{1,2,3} because the translational symmetry implies the conservation of energy-momentum tensor, which was supposed to be the source of the gravitational field.
But most previous formulations were based on curved space-time and, hence, the problems of quantization and energy-momentum conservation remain unsolved.  The general coordinate invariance in Einstein's gravity embodies the local space-time translational symmetry.  Such an invariance corresponds to the group of all transformations of coordinates and is a Lie group with a continuously infinite number of generators.\cite{4}    Noether said: `` Theorem II, finally, in terms of group theory, furnishes the proof of a related Hilbertian assertion about the failure of laws of conservation of energy proper in `general relativity'."  Thus it is natural and desirable that one explores the implication of such a space-time translational symmetry in a conceptual framework similar to that of the Yang-Mills framework.

Recently, it has been shown that there is a consistent and viable formalism of gravity with translational gauge 
symmetry (or T(4) group) in 
flat 4-dimensional space-time based on a generalized Yang-Mills framework.  It leads to an `effective Riemannian metric tensor' in the limit of geometric optics of wave field equations.\cite{3}  
Such a limiting effective metric tensor emerges in the 
Hamilton-Jacobi (or Einstein-Grossmann) equation\footnote{In Yang-Mills gravity based on flat space-time, it may be fitting to call such an equation `Einstein-Grossmann equation' for classical objects, since its form is the same as that obtained in curved space-time for gravity explored by Einstein and Grossmann.\cite{5}} for light rays and classical particles.  In this sense,
the curvature of space-time associated with the classical tests of gravity
 may be interpreted as a classical manifestation of gauge 
fields with translation symmetry in flat space-time.  
Thus, it appears as if light rays and classical particles
in Yang-Mills gravity move in a `curved space-time with 
Riemannian geometry.'  However, the real underlying physical space-time of 
gauge fields and quantum particles 
is flat (i.e.,  vanishing Riemann-Christoffel curvature tensor).
This property of T(4) gauge field in the limit of
geometric optics is essential for Yang-Mills gravity to be  
consistent with all known experiments.\cite{3}  Furthermore, the 
framework of flat space-time enables us to quantize Yang-Mills gravity with a 
well-defined and conserved energy-momentum tensor,
just as the usual gauge theory.    Furthermore, the 
graviton coupling in 
Yang-Mills gravity turns out to be much more simpler than that in Einstein gravity. 
These results and interesting  properties  motivate further investigation 
of the Yang-Mills gravity.  

In sharp contrast to the 
electrodynamics with Abelian group U(1), although the Yang-Mills gravity
is also based on Abelian group T(4) of space-time translation 
symmetry, it  needs Feynman-DeWitt-Mandelstam (FDM) ghosts particles to preserve the gauge invariance and unitarity of the S-matrix. 
The situation is similar to Yang-Mills theories with non-Abelian gauge groups.
Yang-Mills gravity appears to be a natural 
generalization of the conserved charge associated with 
the U(1) group to the conserved energy-momentum tensor 
of the space-time translation group. 
Furthermore, it has a big difference from the usual gauge 
theories with internal gauge groups. Namely, 
the T(4) gauge field in Yang-Mills gravity is not a (Lorentz) vector 
field with dimensionless coupling constant.  Rather, the T(4) gauge field 
is a symmetric tensor field, $\phi_{\mu\nu}=\phi_{\nu\mu}$,
whose coupling 
constant $g$ has the dimension of length (in natural units, $c= \hbar = 
1$).  This is due to the fact that the generators 
$i\p/\p x^{\mu}$ of 
the space-time translation group has the dimension of 1/length and 
cannot be represented by dimensionless constant matrices.  In light of these properties and the unexpected result in Noether's Theorem II regarding general relativity, one cannot take the mathematical and physical properties in usual gauge theories for granted in Yang-Mills gravity.

Furthermore,  we stress that the local space-time translations, $x^\mu \to  x'^{\mu}= x^\mu + \Ld^{\mu}(x)$, where $\Ld^{\mu}(x)$ is an arbitrary and infinitesimal vector function, turns out to be also    the most general infinitesimal coordinate transformation.  As a result, the mathematical and physical contents of such local translations in flat space-time  turns out to be enormously richer and much more difficult to comprehend than the global translations with constant $\Ld^\mu$.
The tensor gauge field  $\phi_{\mu\nu}$ appears to be uniquely associated with the local space-time translation and, simultaneously, the  most general coordinates transformation in flat space-time.  Thus, such a gauge field $\phi_{\mu\nu}(x)$ may be termed `space-time gauge field.'

\section{Translational gauge-invariant action and \\ gauge-fixing 
Lagrangian}
\noindent

Yang-Mills gravity can be formulated in both inertial and 
non-inertial frames and in the presence of fermion 
fields.\cite{3}   It is difficult to discuss quantum field 
theory and particle physics even in a simple non-inertial frame with a 
constant linear acceleration, where the accelerated transformation of 
space-time is 
smoothly connected to the Lorentz transformation in the limit of zero 
acceleration.\cite{6,7,8}  For simplicity, let us consider pure 
quantum Yang-Mills gravity in inertial frames with the Minkowski 
metric tensor $\eta^{\mu\nu}=(1,-1,-1,-1)$.  The  
 action $S_{pg}$ for pure gravity, involving space-time gauge 
 fields $\phi_{\mu\nu}(x)$ 
and a gauge-fixing Lagrangian, is assumed to be\cite{1}
\be
S_{pg}=\int (L_{\phi} + L_{\xi}) d^{4}x,
\ee
\be
L_{\phi}= \frac{1}{4g^2}\left (C_{\mu\nu\a}C^{\mu\nu\a}- 
2C_{\mu\a}^{ \ \ \  \a}C^{\mu\b}_{ \ \ \  \b} \right), 
\ee
$$C^{\mu\nu\a}= J^{\mu\s}\p_{\s} J^{\nu\a}-J^{\nu\s} 
\p_{\s} J^{\mu\a}, \ \ \ \ \  J_{\mu\nu}=\eta_{\mu\nu}+ 
g \phi_{\mu\nu} = J_{\nu\mu} , $$
where $C^{\mu\a\b}$ is the T(4)  gauge curvature and $c=\hbar=1$.  
We note that the Lagrangian
 $L_{\phi}$ changes only by a divergence under the translation gauge 
 transformation, and the action 
functional $S_{\phi}=\int L_{\phi} d^{4}x$ is 
invariant under the space-time translation 
gauge transformation.\cite{1}  To quantize Yang-Mills gravity, it is
necessary to include a gauge fixing 
Lagrangian $L_{\xi}$ in the action functional (1).  For example, the 
gauge fixing Lagrangian enables us to have a well-defined graviton 
propagator (see eq. (33) below).
 The gauge-fixing Lagrangian 
$L_{\xi}$ is assumed to be 
\be
L_{\xi}=\frac{\xi}{2g^{2}}\left[\p^\mu J_{\mu\a} - 
\frac{1}{2} \p_\a J^{\ld}_\ld \right]\left[\p_\nu J^{\nu\a} - 
\frac{1}{2} \p^\a J^{\ld}_\ld \right],
\ee
\be
J^{\ld}_\ld=\delta^{\ld}_{\ld}- g\phi^{\ld}_\ld,  
\ee
where $L_{\xi}$ involves an arbitrary gauge parameter 
$\xi$.  The Lagrangian in (3) corresponds to
a class of gauge conditions of the following form,
\be
\frac{1}{2}[\e^{\mu\rho}\e^{\nu\ld}+\e^{\nu\rho}\e^{\mu\ld} 
-\e^{\mu\nu}\e^{\rho\ld}]
\p_{\ld}J_{\mu\nu}=\p_{\ld} J^{\rho\ld} - \frac{1}{2} \p^{\rho} J^{\ld}_\ld 
= Y^{\rho}, 
\ee
where $Y^{\rho}$ is a suitable function of space-time.

The Lagrangian for pure gravity $L_{pg}=L_{\phi}+
L_{\xi}$ can be expressed in terms of space-time gauge fields $\phi_{\mu\nu}$:
\be
L_{pg}= L_{2}+L_{3} +L_{4} +L_{\xi},
\ee
where
\be
L_{2}=\frac{1}{2}\left(\p_{\ld} \phi_{\a\b} \p^{\ld} \phi^{\a\b}\right.-
\p_{\ld} \phi_{\a\b} \p^{\a} \phi^{\ld\b}-
\p_{\ld} \phi^\b_\b \p^{\ld} \phi^{\s}_\s \ \ \ \ \ \ \ \ 
\ \ \ \ \
\ee
$$+2\p_{\ld} \phi^{\s}_\s \p^{\b} \phi^{\ld}_{\b}- 
\p_{\ld} \phi^{\ld\mu} \p^{\b} \phi_{\mu\b}),$$
 
\be
L_{\xi}=\frac{\xi}{2}\left[(\p_{\ld}\phi^{\ld\a})\p^{\rh}\phi_{\rh\a}
- (\p_{\ld}\phi^{\ld\a})\p_{\a}\phi^{\s}_\s +
\frac{1}{4}(\p^{\a} \phi^\b_\b)\p_{\a} \phi^{\s}_\s \right].
\ee
The Lagrangians $L_{2}$ and $L_{\xi}$ involve quadratic tensor 
field and determine the propagator of the 
graviton in Yang-Mills gravity.  The Lagrangians $L_{3}$ and $L_{4}$ correspond 
to the interactions of 
3- and 4-gravitons respectively.  They can also be obtained from 
the Lagrangian (2).

\section{Vacuum-to-vacuum amplitudes in pure \\ Yang-Mills 
gravity}
\noindent

To quantize a field with gauge symmetry in a covariant formulation, 
one has to impose a gauge condition. In Yang-Mills gravity, it is 
non-trivial to impose a gauge condition in general because the gauge 
condition does not hold for all time.  We know that if one imposes a 
gauge condition in quantum electrodynamics (QED), the gauge condition 
satisfies a free field equation and, hence, hold for all times.  
However, this is true if and only if the gauge condition is linear.  
We have examined the problem of unitarity in QED if we imposed 
a quadratic gauge condition, we found that the gauge 
condition does not hold for all times.\cite{9}  Roughly speaking, the 
longitudinal and time-like photons are no longer free particles, their 
interaction in the intermediate steps of a physical process will 
create extra unwanted amplitudes to upset gauge invariance and 
unitarity of the S-matrix in QED.  Similar to the approach of Faddeev 
and Popov,\cite{10,11} QED with a quadratic gauge condition can be 
described by a total Lagrangian which involves  
ghost particles.  The ghost particles produce extra amplitudes to 
cancel those of unphysical (longitudinal and time-like) photons, so 
that the gauge invariance and unitarity of S-matrix in QED are 
restored.\cite{9}
Similar mechanism of cancellation occurs in any theory with gauge 
symmetry or distorted gauge symmetry, and in Yang-Mills 
gravity.\cite{11,12,13}

We follow Faddeev and Popov's approach to discuss how to fix a gauge for 
all times with the help of path integrals and derived 
the effective Lagrangian for quantum Yang-Mills 
gravity.\cite{10,11}  From the action (1) with the Lagrangian (2) and the 
gauge-fixing terms (3), we 
derived the Yang-Mills field equation
\be
H^{\mu\nu} + \xi A^{\mu\nu} = 0,
\ee
$$
H^{\mu\nu} \equiv \left[\frac{}{}\p_{\ld} (J^{\ld}_{\rho} C^{\rho\mu\nu} - J^{\ld}_{\a} 
C^{\a\b}_{ \ \ \ \b}\eta^{\mu\nu} + C^{\mu\b}_{ \ \ \ \b} J^{\nu\ld}) \right.
$$
\be
\left. - C^{\mu\a\b}\p^{\nu} J_{\a\b} + C^{\mu\b}_{ \ \ \ \b} \p^{\nu} J^{\a}_{\a} -
 C^{\ld\b}_{ \ \ \ \b}\p^{\nu} J^{\mu}_{\ld}\frac{}{}\right]_{(\mu\nu)},
 \ee
\be
A^{\mu\nu} = \left[ \p^{\mu}\left(\p_{\ld} J^{\ld\nu} - \frac{1}{2} \p^{\nu}J^{\ld}_\ld  \right)
- \frac{1}{2}\eta^{\mu\nu}\p^{\ld}\left(\p^{\s} J_{\s\ld} - \frac{1}{2} \p_{\ld}J^{\s}_\s \right)\right]_{(\mu\nu)},
\ee
where $[...]_{(\mu\nu)}$ denotes that $\mu$ and $\nu$ in $[...]$ 
should be made symmetric.  The two terms in (11) are 
gauge-fixing terms, which are non-invariant under gauge 
transformations, similar to that in Einstein gravity.\cite{14}.

Let us consider a general class of the gauge conditions given in 
(5), where $Y^{\a}(x)$ is independent of the 
fields and the gauge function $\Ld^{\a}(x)$.\cite{1}  With such a gauge 
condition, the vacuum-to-vacuum amplitude of the pure Yang-Mills 
gravity is given by
$$
W_Y [j] = \int d[J_{\rh\s}] exp\left(i \int d^{4}x 
(L_{\phi}+ J_{\mu\nu}j^{\mu\nu}) \right)
$$
\be
\times  \ det U \ \prod_{\a,x}
\de(\p^{\ld} J_{\ld\a} - \frac{1}{2} \p_{\a} J^{\ld}_{\ld} - Y_{\a}), 
\ee
where $j_{\mu\nu}$ are external sources. 
The delta function $\de(\p^{\ld} J_{\ld\a} - \frac{1}{2} \p_{\a} J^{\ld}_{\ld} - Y_{\a})$ 
in the path integral (12) is to maintain the gauge 
condition for all times.\cite{12,14}  The functional determinant $det U$ in (12)
is defined by\cite{11,12}
\be
\frac{1}{det U}
= \int d[\Ld^{\rh}(x)]  \ \prod_{x,\a} \de\left(\p^{\ld} J^{\$}_{\ld\a}(x) - \frac{1}{2} \p_{\a} J^{\ld\$}_{\ld} (x) -Y_{\a}(x)\right), 
\ee
\be
J^{\$}_{\mu\nu}=J_{\mu\nu}- \Ld^{\ld}\p_{\ld} J_{\mu\nu} 
-J_{\ld\nu}\p_{\mu} \Ld^{\ld} - J_{\mu\ld}\p_{\nu} \Ld^{\ld},
\ee
where $J_{\mu\nu}^{\$}$ denotes the T(4) gauge transformations of $J_{\mu\nu}$.\cite{1}
The matrix $U$ is obtained by considering the T(4) gauge transformation of 
the gauge condition $\p^{\ld} J_{\ld\a} - \frac{1}{2} \p_{\a}J^{\ld}_\ld = Y_{\a}$.\cite{13}

The matrix U in (13) can be obtained by expressing the Lagrangian for the Feynman-DeWitt-Mandelstam (FDM) ghost fields $\ol{V}^\mu$ and $V^\nu$ in the following form, 
\be
L_{gho}=\ol{V}^{\mu}U_{\mu\nu}V^{\nu},
\ee
where we have 
\be
U_{\mu\nu}= (\p^\ld E_{\mu\nu\ld}) + E_{\mu\nu\ld} \p^\ld
\ee
\be
E_{\mu\nu\ld} = J_{\mu\nu}\p_{\ld} + J_{\nu\ld} \p_\mu + (\p_\nu J_{\mu\ld}) -\e_{\mu\ld} J_{\nu\s} \p^\s -\frac{1}{2}\e_{\mu\ld}(\p_{\nu} J^\s_{\s}).   
\ee                                                                       
Since the vacuum-to-vacuum amplitude $W_Y [j]$ in (12) of Yang-Mills gravity is invariant under an infinitesimal change of 
$Y^{\a}(x)$ for all $Y^{\a}(x)$,\cite{12} we may write 
$W_Y [j]$ in (12) as
\be
W[j]= \int W_Y [j] exp\left[i\int d^{4}x \frac{\xi}{2g^{2}}Y^{\a}(x) Y_{\a}(x) \right] d[Y^{\a}(x)]
\ee
\be 
= \int d[J_{\s\rh}](det \ U) exp \left\{i\int d^{4}x 
\left[\frac{}{}L_{\phi}+\frac{}{}J_{\mu\nu}j^{\mu\nu} + \frac{\xi}{2g^{2}} F_{\a} F_{\b} \e^{\a\b}\right]\right\}
\ee
\be
F_\a \equiv \e^{\rh\ld} \p_{\rh} J_{\ld\a} - \frac{1}{2}\e^{\ld\rh} \p_{\a}J_{\ld\rh},
\ee
\be
det \ U = exp (\ Tr \ ln \ U),
\ee
to within unimportant multiplicative factors.  The amplitude $W[j]$ is 
equivalent to the following total Lagrangian\cite{9,12} of Yang-Mills gravity,
\be
L_{tot}=L_{\phi} + \frac{\xi}{2g^{2}} F_\a F_\b \eta^{\a\b} + 
L_{gho},
\ee
where the FDM ghost Lagrangian $L_{gho}$ is given by (15), (16) and (17).

This total Lagrangian $L_{tot}$ completely specifies the quantum 
Yang-Mills gravity, including the physical tensor gauge field, 
together with the unphysical vector-field $\p^{\mu}\phi_{\mu\nu}$.
and the FDM ghost fields $\ol{V}^{\mu}(x)$ and $V^{\mu}(x)$.  
Thus, the rules for Feynman diagrams in quantum Yang-Mills gravity can be derived from 
the total Lagrangian (22).\cite{13}

\section{Unitarity of the S-matrix and FDM ghost particles}
\bigskip

Let us give some arguments and a proof for the unitarity of 
the S-matrix for Yang-Mills gravity based on the total Lagrangian (22), similar to those of Fradkin and Tyutin for Einstein's gravity.\cite{14}  
For a discussions of unitarity of the Yang-Mills gravity with the 
gauge condition in (5), one can write 
the FDM ghost field $V^{\mu}(x)$ in the following form,\cite{14} 
\be
V^{\mu}(x) = \int d^{4}y {\bf D}^{\mu}_{\nu}(x,y,\phi_{\a\b})\hat{V}^{\nu}(y).
\ee
The FDM ghost field $V^{\mu}$ satisfies the equation
\be
 U_{\mu\nu} V^{\nu} = 0, 
\ee
where $U_{\mu\nu}$ is given in (16) and (17).   Clearly, in the limit of zero coupling strength, $g \to 
0$, one has $J_{\mu\nu} \to \eta_{\mu\nu}$.  Thus, the operator
$U_{\mu\nu}$ reduces to a non-singular differential operator in this 
limit,
\be
U_{\mu\nu} \to  \eta_{\mu\nu}\p_{\ld}\p^{\ld} \equiv 
U^{0}_{\mu\ld}. 
\ee
This limiting property can be seen 
from (16) and (17).  One can choose the function
${\bf D}^{\mu}_{\nu}(x,y,\phi_{\a\b})$ in equation (23) to have the specific 
form
\be
 {\bf D}^{\mu}_{\nu} = \left[U^{-1}\right]^{\mu\ld} \overleftarrow{U}^{0}_{\ld\nu},
\ee
so that $\hat{V}^{\mu}$ satisfies the free field equation,
\be
 U^{0}_{\ld\mu}\hat{V}^{\mu}=[ \p^{\s}\p_{\s} \eta_{\ld\mu} 
]\hat{V}^{\mu}= 0. 
\ee

The generating functional for connected Green's functions in gauge invariant gravity can be 
defined after the gauge condition is specified.\cite{13} 
Similarly, in Yang-Mills gravity the generating functional for 
connected Green's functions (or the 
vacuum-to-vacuum amplitude) (19) can be written as\cite{12}
\be
W[j]= \int d[J_{\a\b}]
exp\left[i\int d^{4}x\left(L_{\phi}+ 
\frac{\xi}{2g^{2}}F_{\mu}F_{\nu}\e^{\mu\nu} + 
J_{\mu\nu}j^{\mu\nu}\right)\right.
\ee
$$
\left. +  Tr \ ln \ U(\overleftarrow{U}^{0})^{-1}\right],  
$$
where the external sources $j^{\mu\nu}$ are arbitrary functions and
$F_{\mu}$ is defined by equation (20).  It follows from (27) and (28) 
that the S-matrix corresponding to the generating functional (28) is 
unitary.\cite{14,12}  The T(4) gauge field equations, $H^{\mu\nu}=0$, 
for pure gravity hold in the physical subspace.

 For a more intuitive and physical understanding of (23) -(28), let us use QED with a non-linear gauge condition,
$$
\p_\mu A^\mu - \b' A_\mu A^\mu = a(x), \ \ \ \ \ a(x):  a \ suitable \ function.
$$
to explain the unitarity of (28) in our argument.  Roughly speaking, the unitarity of the S matrix in QED depends on the property that the unphysical components, $ \chi=\p_\mu A^\mu$, of the gauge field $A^\mu$ satisfies the free equation
$
\p_\mu \p^\mu \chi = 0.
$
This is indeed the case if one impose a linear gauge condition in QED for quantization.  Now suppose one imposes the non-linear gauge condition instead of the usual linear gauge condition.  One can show that $ \chi=\p_\mu A^\mu$ no longer satisfies the free equation.  Instead, one has the equation$
(\p_\mu -2\b' A_\mu)\p^\mu a(x) \equiv M a(x) = 0$.
  This equation implies that the unphysical components of $A^\mu$ have a new interaction with $A_\mu$ (in other words, the non-linear gauge condition (1) cannot be imposed for all times).\cite{9}  Such interaction will produce extra unphysical amplitudes to violate unitarity (and gauge invariance) of the S matrix.  Therefore, these unphysical amplitudes must be removed from the theory to restore unitarity of QED.  In gauge theory, this is accomplished by the presence of 
$det M= exp[Tr  \ ln \ M] $
in the generating functional of Green's function.  In other words, the non-free operator $M$ is 'subtracted' from QED.  Then there is no  more unitarity-violation amplitudes in QED.  Effectively, the unphysical components of the photon satisfies the free equation, so that  QED with non-linear gauge condition becomes unitary again.

The last term in (28) can  be 
written in terms of FDM ghost fields $V^{\a}(x)$ and 
$\ol{V}^{\b}(x)$,\cite{14,15}
\be
exp\left(Tr \ ln  \ U(\overleftarrow{U}^{0})^{-1}\right) = \int 
d[V^{\a},\ol{V}^{\b}]exp\left(i\int L_{gho}d^{4}x\right),
\ee
where the ghost Lagrangian $L_{gho}$ is given by (15), which  
describes the FDM ghosts associated with the gauge specified in 
(5). 
Note that $\ol{V}^{\mu}$ is considered as an independent field.
The quanta of the fields $V^{\mu}$ and $\ol{V}^{\mu}$ in the 
 Lagrangian $L_{gho}$ are the FDM ghost 
particles, which are vector-fermions.  By definition of the physical 
states for the S-matrix, these FDM ghost particles do not exist in the 
external states.  They can only appear in the intermediate 
steps of a physical process.\cite{12}  

\section{Graviton self-energy}

Yang-Mills gravity with space-time translational gauge symmetry has interesting properties.  On one hand, the graviton self-coupling has a maximum of four gravitons at a vertex, which resembles to that in renormalizable gauge theories such as QCD and electroweak theory.  On the other hand,  the total Lagrangian (22) appears to be not renormalizable by naive power counting, because the coupling strength $g$ has the dimension of length, contrary to usual gauge theories.  In order to shed some light on the situation, we consider the graviton self-energy at the one-loop level and compare the result in Yang-Mills gravity with other theories.  We employ the D-dimensional regularization to preserve the space-time translational gauge symmetry of physical amplitudes and to define the ultraviolet divergent quantities. 

To explore renormalization at the one-loop level, let us consider the divergent self-energy of the graviton, which is contributed from several Feynman diagrams.  The most important contribution came from the diagram corresponding to the process,
\be
\phi_{\mu\nu}(p ) \to \phi_{\rho\s}(q) \phi_{\g\ld}(p-q) \to \phi_{\a\b}(p ).
\ee
Its amplitude, denoted by $S_{1}^{\mu\nu\a\b}(p )$, can be calculated by using the Feynman rules with $\xi =2$ in the gauge-fixing Lagrangian (3), for simplicity.\cite{13}   The process (30) leads to the following amplitude for the graviton self-energy,
\be
S_{1}^{\mu\nu\a\b}=\int \frac{d^{D}q}{(2\pi)^D} V^{\mu\nu\rho\s\g\ld}(p,q,k) G_{\rho\s\rho'\s'}(q)
\ee
$$
\times G_{\g\ld\tau\eta}(k_1)V^{\a\b\rho'\s'\tau\eta}(p_1,q_1,k_1), 
$$
where  $ D \to 4$ at the end of calculations.  Since all momenta are incoming to the vertex, by definition, we have $p+q+k=0, q_{1}=-q, k_{1}=-k =p+q, p_{1}=-p.$  The graviton 3-vertex,  $[\phi^{\mu\nu}(p)\phi^{\s\tau}(q)\phi^{\ld\rh}(k)] \equiv V^{\mu\nu\s\tau\ld\rh}(p,q,k)$, is given by\cite{13}
\be
ig \ Sym \ P_{6}\left(\frac{}{}-p^{\ld}q^{\rho}\e^{\s\mu}\e^{\tau\nu}
+p^{\s}q^{\rho}\e^{\ld\mu}\e^{\tau\nu}+p^{\ld}q^{\rh}\e^{\s\tau}\e^{\mu\nu}\right.
\ee
$$ 
\left.-p^{\s}q^{\rho}\e^{\mu\nu}\e^{\tau\ld} 
 -p^{\rho}q^{\s}\e^{\mu\nu}\e^{\tau\ld}
+p^{\mu}q^{\rho}\e^{\nu\s}\e^{\ld\tau}\frac{}{}\right).$$
We use the symbol Sym to denote a symmetrization is to be performed on each index 
pair $(\mu\nu)$, $(\s\tau)$ and $(\ld\rh)$  in $[\phi^{\mu\nu}(p)\phi^{\s\tau}(q)\phi^{\ld\rh}(k)]$.  
The symbol  $P_{n}$ denotes a summation is 
to be carried out over permutations of the momentum-index 
triplets, and the subscript gives the number of permutations 
in each case.   The graviton propagator for gauge parameters $\zeta=1$ and arbitrary $\xi$ is
\be
G_{\a\b\rh\s}(k)=-i\left[\frac{1}{2 k^{2}}(\e_{\a\b}\e_{\rh\s}-
\e_{\rh\a}\e_{\s\b}-\e_{\rh\b}\e_{\s\a})\right.
\ee
$$
+ \left.\frac{1}{k^{4}}\frac{\xi-2}{2\xi}(k_{\s}k_{\b}\e_{\rh\a}
+k_{\a}k_{\s}\e_{\rh\b}+ k_{\rh}k_{\b}\e_{\s\a}+
k_{\rh}k_{\a}\e_{\s\b})\right],$$
where the $i\ep$ prescription for the Feynman propagators is understood.   In the limit, $D \to 4$, FeynCalc\cite{18} leads to the self-energy (31) due to the graviton loop,
\be 
S_{1}^{\mu\nu\a\b} =\frac{i g^2}{24(4\pi)^2} \G(0)[\frac{}{} p^4(8\e^{\a\nu}\e^{\b\mu}+8\e^{\a\mu}\e^{\b\nu} + 31\e^{\a\b}\e^{\mu\nu}) 
\ee
$$
+ 56 p^\a p^\b p^{\mu} p^{\nu} -30p^2 (p^\a p^\b \e^{\mu\nu} + p^\mu p^\nu \e^{\a\b})
$$
$$
 -7p^2 (p^\a p^\mu \e^{\b\nu}+p^\b p^\mu \e^{\a\nu}+p^\a p^\nu \e^{\b\mu}+p^\b p^\nu \e^{\a\mu}),
$$
where we have set $\xi=2$ for simplicity, $p^4=(p^2)^2=(p_{\mu}p^{\mu})^2$ and $\G(0)$ is the Euler function $\G(n)$ with $n=0$.  The factor $1/2$ for the Feynman diagram of the process (30) is included in (34).  This result (34) is consistent with the symbolic computing using Xact.\cite{18}

There is a graviton self-energy diagram involving FDM ghost-loop, which corresponds to the process which may be denoted by
\be
\phi_{\mu\nu}(p ) \to V_{\rho}(q) \ol{V}_{\ld}(p-q) \to \phi_{\a\b}(p ).
\ee
where  the vector ghost-loop is associated with a factor $(-1)$ due to the fermion property of the vector ghost field.  The virtual process (35) leads to the self-energy amplitude,
\be
S_{2}^{\mu\nu\a\b} (p )=\int \frac{d^{D}q}{(2\pi)^D} V^{\mu\nu\rho\ld}(p,q,k) G_{\rho\rho'}(q)
\ee
$$
\times G_{\ld\ld'}(k)V^{\a\b\rho'\ld'}(p_1,q_1,k_1).
$$
The graviton-ghost-ghost vertex $\phi^{\mu\nu}(p) \ol{V}^{\rh} (q) V^{\ld} (k)=V^{\mu\nu\rh\ld} (p,q,k)$ and the ghost propagator $G^{\mu\nu}(q)$ are given by\cite{13}
\be
V^{\mu\nu\rh\ld} (p,q,k)=\frac{ig}{2} [q^\mu p^\ld \e^{\rh\nu}+q^\nu p^\ld \e^{\rho\mu} +(k \cdot q)(\e^{\rh\mu}\e^{\ld\nu} +\e^{\rh\nu}\e^{\ld\mu} )
\ee
$$
+q^\mu k^\rh \e^{\ld\nu}+ q^\nu k^\rh \e^{\ld\mu}-q^\rh p^\ld \e^{\mu\nu}-q^\rh k^\nu \e^{\ld\mu}-q^\rh k^\mu \e^{\ld\nu}],
$$
and 
\be
G^{\mu\nu} (q) = \frac{-i}{q^2} \e^{\mu\nu}.
\ee
FeynCalc leads to the result
\be
S_{2}^{\mu\nu\a\b} = \frac{-i}{48}\frac{g^2}{(4\pi)^2}
\G(0) p^2 [ 7p^4 \e^{\a\b}\e^{\mu\nu}+  2p^4 (\e^{\a\mu}\e^{\b\nu} +\e^{\a\nu}\e^{\b\mu})
\ee
$$
- p^2 (p^\a p^\mu \e^{\b\nu}+ p^\b p^\mu \e^{\a\nu}+ p^\a p^\nu \e^{\b\mu}+ p^\b p^\nu \e^{\a\mu}) .
$$
$$
-6 p^2 (p^{\a}p^{\b} \e^{\mu\nu} + p^\mu p^{\nu}\e^{\a\b}) + 8p^{\mu}p^{\nu}p^{\a}p^{\b}].
$$

At the one-loop level, there are three more tadpole diagrams involving (i) a graviton-loop and a graviton 4-vertex, (ii) a graviton-loop and two graviton 3-vertices, and (iii) a ghost-loop with one graviton 3-vertex and one graviton-ghost-ghost vertex.  When we use dimensional regularization to calculate the diagram (i) with a factor 1/2, we have
\be
S_{3}^{\mu\nu\a\b}=\frac{1}{2} \int \frac{d^D q}{(2\pi)^D} V^{\mu\nu\s\tau\rh\ld\a\b}(p,q,k,l) G_{\s\tau\ld\rh}(q),
\ee
where the 4-vertex $V^{\mu\nu\s\tau\rh\ld\a\b}(p,q,k,l)$ involves terms of the form\cite{13}
\be
p^\rh q^\b \e^{\mu\s}\e^{\nu\tau}\e^{\ld\a}, etc.
\ee
Each term has one momentum $q^\b$ and a D-dimensional integration, so that it vanishes,
\be
\int \frac{d^D q}{(2\pi)^D} q^\b f(q^2) = 0,
\ee
because it is an odd function in $q^\a$.   The other two tadpole diagrams (ii) and (iii) also vanish under dimensional regularization due to (42) and
\be
\int \frac{d^D q}{(2\pi)^D} (q^2)^{\b-1} = 0, \ \ \ \ \   \b=0,1,2,3,É.
\ee
Note that these two tadpole diagrams have additional complication.  Strictly speaking, they are undefined because they take  the form of zero divided by zero.  The reason is that the loop and the graviton line, $\phi_{\mu\nu}(p ) \to \phi_{\a\b}(p')$ are connected by a graviton propagator involving a factor $1/(p-p')^2$ which diverges in the limit $p \to p'$.\cite{19}  Here, in consistent with the spirit of the dimensional regularization, we impose a rule that    the dimensional regularization is always taken first before other limiting procedure is carried out.  Equivalently, one may temporarily introduce a small mass for the graviton to avoid infinity and  then one takes the zero mass limit after the dimensional regularization is carried out.
In gauge theories with internal gauge symmetry, one might include counter terms to remove the contributions from these tadpole diagrams.\cite{12}

The total divergent self-energy amplitudes $S^{\mu\nu\a\b}(p )$ is the sum of (34) and (39), i.e.,
\be
S^{\mu\nu\a\b}=S_{1}^{\mu\nu\a\b}+S_{2}^{\mu\nu\a\b}= \frac{i g^2}{48(4\pi)^2} \G(0) 
\ee
$$
 \times \left[\frac{}{}p^4(14\e^{\a\nu}\e^{\b\mu}+14\e^{\a\mu}\e^{\b\nu} + 55 \e^{\a\b}\e^{\mu\nu})\right. 
$$
$$
+ 104 p^\a p^\b p^{\mu} p^{\nu} - 54p^2 (p^\a p^\b \e^{\mu\nu} + p^\mu p^\nu \e^{\a\b})
$$
$$
 -\left.  13 p^2 (p^\a p^\mu \e^{\b\nu}+p^\b p^\mu \e^{\a\nu}+p^\a p^\nu \e^{\b\mu}+p^\b p^\nu \e^{\a\mu})\frac{}{}\right].
$$

\section{Discussions}

\noindent
The graviton propagator is modified because the graviton can virtually disintegrated into two gravitons and an FDM ghost pair for a small fraction of time, as indicated in (30) and (35) respectively.  The ultraviolet divergence of the graviton self-energy contains only the simple pole terms in (44) after dimensional regularization.  In this sense, the ultraviolet divergence in Yang-Mills gravity is no worse than that in QED at the one-loop level.  Just like massless gauge theory, there is another type of divergence at low energies, i.e., the infrared divergence, in Yang-Mills gravity.  We shall not consider it in the paper.

However, in QED it is easy to see how such a self-interaction affect the photon self energy by summing up the free photon propagator and its one-loop correction.\cite{20}  In contrast, the situation is much more complicated for the graviton.  For simplicity, let us consider the graviton propagator (33) with $\xi=2$.  We have 
\be
G_{\a\b\rh\s}(p)=-i\left[\frac{1}{2 p^{2}}(\e_{\a\b}\e_{\rh\s}-
\e_{\rh\a}\e_{\s\b}-\e_{\rh\b}\e_{\s\a})\right],
\ee
which corresponds to the Feynman gauge in QED.   One calculates the one-loop correction of $G_{\mu\nu\a\b}$, i.e.,
\be
G^{(1)}_{\mu\nu\a\b}=G_{\mu\nu\mu'\nu'} S^{\mu'\nu'\a'\b'} G_{\a'\b'\a\b},
\ee
by using (44) and (45).   We obtain
\be
G^{(1)}_{\mu\nu\a\b}=\frac{- i g^2 \G(0) }{48(4\pi)^2 p^4} \left[\frac{}{}14 p^4(\e_{\a\nu}\e_{\b\mu}+\e_{\a\mu}\e_{\b\nu} +  \e_{\a\b}\e_{\mu\nu})\right. 
\ee
$$
+ 104 p_\a p_\b p_{\mu} p_{\nu} + 28p^2 (p_\a p_\b \e_{\mu\nu} + p_\mu p_\nu \e_{\a\b})
$$
$$
 -\left. 13 p^2 (p_\a p_\mu \e_{\b\nu}+p_\b p_\mu \e_{\a\nu}+p_\a p_\nu \e_{\b\mu}+p_\b p_\nu \e_{\a\mu})\frac{}{}\right].
 $$
We see that the form (47) differs from that of a free graviton propagator (45).  We have checked these properties carefully.  This difference in the structure between (47) and (45) appears to suggest that  quantum corrections to graviton self-energy in perturbation does not imply the mass of graviton in Yang-Mills gravity, in sharp contrast to QED.  This difference may be related to the fact that Yang-Mills gravity is based on external  space-time translational gauge symmetry, while QED is based on the internal $U_1$ gauge symmetry.
It also suggests that the structure of counter terms to remove the divergent self-energy of gravitons is more complicated than that in QED, as expected for a tensor gauge field.  The graviton self-energy has complicated dependence of gauge parameters, as one can see in equations (A.3), (A.11) and (A.12) in the  appendix.  However, the divergent terms at the one-loop level in Yang-Mills gravity can be removed by a gauge-dependent counter term, similar to that in gauge field theories.

 If one compares the gauge dependence of the graviton self-energy (A.12) in the appendix with the results\cite{21,22,23} for graviton self-energy in Einstein's gravity, one sees that they also have similar gauge-dependence at the one-loop level.  These divergent quantities are not directly observable, so their gauge dependence is not a problem.  We note that in Einstein's gravity, the gravitons 
have N-vertex of self-coupling, where N is an arbitrarily
large number, which is in sharp contrast to the maximum 4-vertex  for self-coupling of graviton in Yang-Mills gravity.  Thus, we expect that in the higher order processes when the N-vertex, $N\ge 5$, is involved, the results in Yang-Mills gravity will be simpler than the corresponding results in Einstein's gravity, in which as N increases the number of counter terms will also increase without limit.  Such complication will not occur in Yang-Mills gravity because of the simplicity of the graviton self-coupling, $N \le 4$.  But this property does not necessarily imply a finite number of counter terms in Yang-Mills gravity.

We have given a formal argument for the unitarity of the S-matrix in Yang-Mills gravity with $T_4$ gauge symmetry, in analogy to that in Einstein gravity.  It is desirable that the unitarity of the S matrix is also substantiated by explicit calculations of gravitational scattering processes of physical particles in Yang-Mills gravity.  

The motivation for investigating the high energy behavior  of quantum gravity is to have a complete understanding of the whole of field theory rather than its experimental verifications.  It seems fair to say that the experimental effects of quantum gravity cannot be detected in the earth laboratories, because of the extremely weakness of its coupling strength ($\approx 10^{-39}$).  However, we would like to make sure that observable results in QCD, electroweak theory and all other calculations at a relatively low energy\cite{24,25,26} will not be upset by the higher order corrections due to the gravitational interaction.  This is important theoretically because all physical particles in nature cannot escape from the gravitational interaction.

Furthermore, one would like to know whether the Yang-Mills idea of gauge symmetry is powerful enough to accommodate all interaction forces in nature,  including gravity.  The answer appears to be affirmative.  We have also demonstrated that all strong, electroweak and gravitational interactions can be unified in a model based on the generalized Yang-Mills framework.\cite{27,28} The generalized framework can accommodate both the usual gauge symmetries of compact internal groups and the non-compact translational symmetry group of external space-time.  The idea of such a total unified model resembles Glashow-Salam-Ward-Weinberg's idea for electroweak unification.  This is interesting because the unified model shows that the gravity can be brought back to the framework of Yang-Mills field theories based on flat space-time.  This general symmetry framework based on flat space-time can accommodate    ($\a$) a unification of all interactions in nature and ($\b$) all essential requirements for physics, such as conservation laws, quantizations of fields and physical frames of reference (inertial and non-inertial).\cite{6,7,8}  It was termed `taiji\footnote{In ancient Chinese thought, the word `taiji' denotes the ultimate principle or the condition that existed before the creation of the world.}  symmetry framework'\cite{29} to stress its potential significance for all physics in any physical frame of reference.

  For an ordinary tensor field 
theory with a dimensional coupling constant (such as $g$) and without having a gauge 
symmetry, one would conclude that the theory is not renormalizable 
based on power counting.  Within the generalized Yang-Mills framework, it is almost certain that the dimensional coupling constant $g$ is inherent in the gravitational interaction.  Nevertheless, the usual argument of power counting may not be 
applicable to Yang-Mills gravity with T(4) gauge symmetry and dimensional regularization in flat space-time.\footnote{On the other hand, there is a possibility that
Yang-Mills gravity may be `renormalizable' in the modern sense, i.e.,  every ultraviolet divergence in the theory can be canceled by a counter term.\cite{30,31,32}}  To clarify the situation, one may have to investigate, say, the self-energy of graviton at the 2-loop level or higher.

\bigskip
\bigskip
{\bf Acknowledgements}
\bigskip

The authors would like to thank Dana Fine, Leonardo Hsu, Jay Wang and G. Khanna for discussions.
  The work was supported in part by Jing Shin
Research Fund of the UMass Dartmouth Foundation and by the doctoral fellowship of UMassD.  Some results of symbolic computing (using R. Mertig's `FeynCalc' and J. M. Martin-Garcia's package `xAct') were obtained in the UMass Dartmouth High-Performance Computing Cluster (NSF grant CNS-0959382 and AFOSR DURIP grant FA9550-10-1-0354).

\bigskip

\bigskip

\renewcommand\theequation{{A.1}} 
 \renewcommand\thesection{{A-}\@arabic\c@section} 

\bigskip

\bigskip

{\large \bf   Appendix   \ \ \  Graviton self-energy in general gauge } 

\bigskip
To show the gauge-dependent self-energy of graviton, let us consider a general gauge condition, which is a generalization of the gauge-fixing Lagrangian (3) to involve two parameters, $\xi$ and $\zeta$ :
\be
L_{\xi\zeta}=\frac{\xi}{2g^{2}}\left[\p^\mu J_{\mu\a} - 
\frac{\zeta}{2} \p_\a J^{\ld}_\ld \right]\left[\p_\nu J^{\nu\a} - 
\frac{\zeta}{2} \p^\a J^{\ld}_\ld \right].  
\ee
In this case, the corresponding graviton propagator will also be a generalization of (33) with two parameters,
$$
G_{\a\b\rh\s}(k)=\frac{-i}{2}\left[\frac{}{}\frac{1}{ k^{2}}(\e_{\a\b}\e_{\rh\s}-
\e_{\rh\a}\e_{\s\b}-\e_{\rh\b}\e_{\s\a})\right.
$$
$$
+ \left.\frac{1}{k^{4}}\frac{\xi-2}{\xi}(k_{\s}k_{\b}\e_{\rh\a}
+k_{\a}k_{\s}\e_{\rh\b}+ k_{\rh}k_{\b}\e_{\s\a}+
k_{\rh}k_{\a}\e_{\s\b})\right.,
$$
$$
- \frac{2(\zeta -1)}{k^4 (\zeta -2)}(k_\a k_\b \e_{\rh\s} + k_{\rh} k_{\s} \e_{\a\b})
$$
\renewcommand\theequation{{A.2}} 
\be
\left. - \frac{4(\zeta - 1)}{k^6 \xi (\zeta-2)^2}(6-\xi-2\xi\zeta - 2\zeta)k_\a k_\b k_\rh k_\s \right],
\ee
where the $i\ep$ prescription for the Feynman propagators is understood.   In the limit, $D \to 4$, FeynCalc\cite{18} leads to the self-energy $S_{1}^{\mu\nu\a\b}$ due to the graviton loop,
$$
S_{1}^{\mu\nu\a\b} =\frac{i g^2}{(4\pi)^2}\frac{ \G(0)}{240 \xi^2 (\zeta-2)^4}\left[\frac{}{} S_a  p^4(\e^{\a\nu}\e^{\b\mu}+\e^{\a\mu}\e^{\b\nu}) \right.
$$
$$
+ S_b   p^4  \e^{\a\b}\e^{\mu\nu} + S_c  p^\a p^\b p^{\mu} p^{\nu} + S_d  p^2 (p^\a p^\b \e^{\mu\nu} + p^\mu p^\nu \e^{\a\b})
$$
\renewcommand\theequation{{A.3}} 
\be 
 + S_e  p^2 (p^\a p^\mu \e^{\b\nu}+p^\b p^\mu \e^{\a\nu}+p^\a p^\nu \e^{\b\mu}+p^\b p^\nu \e^{\a\mu})],
\ee
where
\renewcommand\theequation{{A.4}} 
\be
S_a=[(69\xi^2 + 80 \xi +92)\zeta^4 - 8(54\xi^2 +47\xi +92)\zeta^3
\ee
$$
+8(165\xi^2 + 52 \xi +252)\zeta^2 - 32(60\xi^2 -\xi +68)\zeta
$$
$$
+4(273\xi^2 - 60 \xi +196)],
$$
\renewcommand\theequation{{A.5}} 
\be
S_b=8 [(3\xi^2 + 20 \xi +19)\zeta^4 + (101\xi^2 - 62\xi -152)\zeta^3
\ee
$$
+(-425\xi^2 + 112 \xi +312)\zeta^2 + 8(85\xi^2 - 67\xi - 4)\zeta
$$
$$
-381\xi^2 +620 \xi - 212],
$$
\renewcommand\theequation{{A.6}} 
\be
S_c=16 [(17\xi^2 + 10 \xi +36)\zeta^4 - 3 (17\xi^2 + 6\xi + 96)\zeta^3
\ee
$$
+2 (75\xi^2 -56 \xi +404)\zeta^2 - 8(35\xi^2 - 22\xi +116)\zeta
$$
$$
+186\xi^2 - 40 \xi +392],
$$
\renewcommand\theequation{{A.7}} 
\be
S_d=-2 [(17\xi^2 + 60 \xi +76)\zeta^4 +8 (58\xi^2 -11\xi - 76)\zeta^3
\ee
$$
-8 (225\xi^2 +34 \xi - 156)\zeta^2 + 16(155\xi^2 - 44\xi -8)\zeta
$$
$$
-4(301\xi^2 - 380 \xi +212)],
$$
\renewcommand\theequation{{A.8}} 
\be
S_e= [(21\xi^2 - 120 \xi -92)\zeta^4 + (-68\xi^2 +776\xi +736)\zeta^3
\ee
$$
-32 (10\xi^2 +53 \xi +68)\zeta^2 + 16(65\xi^2 +98\xi +176)\zeta
$$
$$
-4(193\xi^2 +100 \xi + 356)],
$$
where we have used (31), (32) and (A.2). 

For the general gauge-fixing Lagrangian (A.1), the graviton-ghost-ghost vertex,
$ \ol{V}^{\mu}( p) V^{\nu}(q)\phi^{\alpha\beta}(k)\equiv U^{\mu\nu\alpha\beta}(p,q,k) $, and the ghost propagator $G^{\mu\nu}(q)$  are respectively given by 
 \renewcommand\theequation{{A.9}} 
\be
U^{\mu\nu\alpha\beta}(p,q,k) = \frac{i g}{2}\left[ p^{\alpha}k^{\nu}\eta^{\mu\beta} + p^{\beta}k^{\nu}\eta^{\alpha\mu}+ p\cdot q \left(\eta^{\mu\alpha}\eta^{\nu\beta} + \eta^{\mu\beta}\eta^{\nu\alpha}\right) \right.  
\ee
$$
 +p^{\alpha} q^{\mu} \eta^{\beta\nu} + p^{\beta}q^{\mu} \eta^{\alpha\nu}  
- \zeta p^{\mu}k^{\nu} \eta^{\alpha\beta} - \zeta p^{\mu}q^{\beta} \eta^{\alpha\nu} 
-\left. \zeta p^{\mu}q^{\alpha} \eta^{\beta\nu} \right],
$$
\renewcommand\theequation{{A.10}} 
\be
G^{\mu\nu}(q) = \frac{-i}{q^2}\left(\eta^{\mu\nu} - \frac{q^{\mu}q^{\nu}}{q^2}\frac{(1-\zeta)}{(2-\zeta)}\right).  
\ee
FeynCalc leads to the following contribution of the ghost-loop to the graviton self-energy,
\renewcommand\theequation{{A.11}} 
\be
S_2^{\mu\nu\alpha\beta}  =  - \frac{i g^2}{(4\pi)^2}\frac{\Gamma(0)}{480\, (\zeta - 2)^2}\left[(15\zeta^2 - 46\zeta + 51) \, p^4\left(\eta^{\alpha\nu}\eta^{\beta\mu}+\eta^{\alpha\mu}\eta^{\beta\nu}\right) \right.
\ee
$$
 \left.+(5\zeta^2 + 14\zeta + 51) \, p^4 \eta^{\alpha\beta}\eta^{\mu\nu} + 8(15\zeta^2 - 46\zeta + 41) \, p^{\alpha}p^{\beta}p^{\mu}p^{\nu}  \right.  
$$
$$
 \left. + 2(18\zeta - 23)\, p^2 \left(p^{\alpha}p^{\mu}\eta^{\beta\nu} + p^{\beta}p^{\mu}\eta^{\alpha\nu}+ p^{\alpha}p^{\nu}\eta^{\beta\mu}+ p^{\beta}p^{\nu}\eta^{\alpha\mu}\right) \right. 
 $$
 $$
  \left.- 2(5\zeta^2 + 22\zeta + 3) \, p^2 \left(p^{\alpha}p^{\beta}\eta^{\mu\nu} + p^{\mu}p^{\nu}\eta^{\alpha\beta}\right) \, \right],
$$
where the ghost-loop is associated with a factor $(-1)$ due to the fermion nature of FDM ghost.  
Finally, we obtain the total self-energy amplitude $T^{\mu\nu\alpha\beta}$ :
\renewcommand\theequation{{A.12}} 
\be
T^{\mu\nu\alpha\beta} = S_1^{\mu\nu\alpha\beta} +S_2^{\mu\nu\alpha\beta} 
\ee
$$
= \frac{i g^2}{(4\pi)^2}\frac{\Gamma (0)}{480\, \xi^2 (\zeta-2)^4 }\left[T_1 \, p^4\left(\eta^{\alpha\nu}\eta^{\beta\mu}+\eta^{\alpha\mu}\eta^{\beta\nu}\right)+T_2 \, p^4 \eta^{\alpha\beta}\eta^{\mu\nu} \right. 
$$
$$ 
+ T_3 \, p^{\alpha}p^{\beta}p^{\mu}p^{\nu}+ T_4 \, p^2 \left(p^{\alpha}p^{\beta}\eta^{\mu\nu} + p^{\mu}p^{\nu}\eta^{\alpha\beta}\right) 
$$
$$
+ \left. T_5 \, p^2 \left(p^{\alpha}p^{\mu}\eta^{\beta\nu} + p^{\beta}p^{\mu}\eta^{\alpha\nu}+ p^{\alpha}p^{\nu}\eta^{\beta\mu}+ p^{\beta}p^{\nu}\eta^{\alpha\mu}\right) \right],
$$
 
where the coefficients  $T_1 ,T_2 ... T_5$ are given by
\renewcommand\theequation{{A.13}} 
 \be
 T_1 =  \left(123 \xi ^2+160 \xi +184\right)\zeta ^4 -2 \left(379 \xi ^2+376 \xi
   +736\right)\zeta ^3   
   \ee
   $$   + \left(2345 \xi ^2+832 \xi +4032\right)\zeta ^2 -4 \left(863 \xi
   ^2-16 \xi +1088\right)\zeta 
   $$ 
   $$
      +4 \left(495 \xi ^2-120 \xi +392\right) .   
      $$
\renewcommand\theequation{{A.14}} 
 \be      
T_2 =  \left(43 \xi ^2+320 \xi +304\right)\zeta ^4 +2 \left(811 \xi ^2-496 \xi
   -1216\right)\zeta ^3 
   \ee
   $$
      -\left(6815 \xi ^2 -1792 \xi -4992\right)\zeta ^2 +4 \left(2757
   \xi ^2-2144 \xi -128\right)\zeta  
   $$
   $$   -6300 \xi ^2+9920 \xi -3392 .   
   $$
   \renewcommand\theequation{{A.15}} 
 \be
T_3 = 8 \left[\left(53 \xi ^2+40 \xi +144\right) \zeta ^4  -2 \left(49 \xi ^2+36 \xi
   +576\right)\zeta ^3 \right. \
   \ee
   $$
   \left. +\left(315 \xi ^2-448 \xi +3232\right)\zeta ^2 -4 \left(193 \xi
   ^2-176 \xi +928\right)\zeta \right. 
   $$
   $$   
   \left. +4 \left(145 \xi ^2-40 \xi +392\right)\right] .    
   $$
\renewcommand\theequation{{A.16}} 
\be
T_4 = -2 \left[\left(29 \xi ^2 +120 \xi +152\right)\zeta ^4 +\left(926 \xi ^2 -176
   \xi -1216\right)\zeta ^3 \right. 
 \ee
 $$
  \left. -\left(3535 \xi ^2+544 \xi -2496\right)\zeta ^2 +\left(4884 \xi ^2 -1408 \xi -256\right)\zeta \right. 
  $$
  $$   
   \left. -4 \left(605 \xi ^2-760 \xi +424\right)\right] . 
   $$
 \renewcommand\theequation{{A.17}} 
 \be
T_5 =  2 \left[ \left(21 \xi ^2-120 \xi -92\right)\zeta ^4 -\left(86 \xi ^2-776 \xi
   -736\right)\zeta ^3  \right. 
   \ee
   $$
    \left.  -\left(225 \xi ^2+1696 \xi +2176\right)\zeta ^2 +4 \left(219 \xi
   ^2+392 \xi +704\right)\zeta \right. 
   $$
   $$
   \left.-8 \left(85 \xi ^2+50 \xi +178\right)\right]  .
$$


\newpage

\bibliographystyle{unsrt}

\begin{thebibliography}{99}
\bibitem{18}We employed R. Mertig's `FeynCalc' and J. M. Martin-Garcia's package `xAct' to carry symbolic computing and to check consistency.  Our program was also used to evaluate the self-energy in Einstein gravity, we obtained consistent results. See ref. 22.,[
\bibitem{22}The `FeynCalc' is also used to check the gauge-dependent self-energy of graviton  in ref. 27.  Our result agrees with Capper's result, except that there are typos and missing terms in the results of gauge-dependent self-energy in ref. 27.  For example, the factors $(\b^2/\a)$ and (1045/16) in equation (2.19) should read $(\b^2/\a^2)$ and (1405/16) respectively.
\end{thebibliography}

\end{document}